\documentclass[12pt,a4paper]{article}

\usepackage{ascmac}
\usepackage[dvipdfmx]{graphicx}
\usepackage{amsmath}
\usepackage{amsfonts}

\def\tr{\mathop{\rm tr}\nolimits}

\def\Pexp{\mathop{\rm Pexp}\nolimits}
\def\Plog{\mathop{\rm Plog}\nolimits}

\newcommand{\fq}{{\mathfrak{q}}}

\begin{document}
\begin{titlepage}
\title{
\vspace{-1.5cm}
\begin{flushright}
{\normalsize TIT/HEP-689\\ May 2022}
\end{flushright}
\vspace{1.5cm}
\LARGE{Analytic continuation for giant gravitons}}
\author{
Yosuke {\scshape Imamura\footnote{E-mail: imamura@phys.titech.ac.jp}}
\\
\\
{\itshape Department of Physics, Tokyo Institute of Technology}, \\ {\itshape Tokyo 152-8551, Japan}}

\date{}
\maketitle
\thispagestyle{empty}

\begin{abstract}
We investigate contributions of giant gravitons
to the superconformal index.
We concentrate on coincident giant gravitons
wrapped around a single cycle,
and each contribution
is
obtained by a certain variable change for fugacities from
the index of the worldvolume theory on the
giant gravitons.
Because we treat the index as a series
of fugacities and
the variable change relates different
convergence regions,
we need an analytic continuation
before summing up such contributions.
We propose a systematic prescription for the continuation.
Although our argument is based on
some unproved assumptions,
it passes non-trivial numerical
checks for some examples.
With the prescription we can calculate the indices of
the M5-brane theories from those of the
M2-brane theories, and vice versa.
\end{abstract}
\end{titlepage}

\tableofcontents

\section{Introduction}
For the last few decades the AdS/CFT correspondence \cite{Maldacena:1997re,Gubser:1998bc,Witten:1998qj} 
has been playing an important role
in the progress of string theory and quantum field theories.
It provides novel approaches for
investigation of different physical quantities in various situations.
It is very powerful for the analysis of large $N$ gauge theories,
and even in the finite $N$ region, where the Planck length is not negligible,
it is possible to calculate
supersymmetry protected quantities
like $R$-charges of gauge invariant operators by using
string theory
in the AdS background.

Generating functions of such quantities have been calculated on the gravity side of the AdS/CFT corresponsence.
For example,
the superconformal index
of the ${\cal N}=4$ $U(N)$ SYM in the large $N$ limit 
was reproduced as the index of the bulk supergravity modes \cite{Kinney:2005ej}.
Extended branes are important when $N$ is finite.
It is known that not only branes wrapped around topologically non-trivial cycles \cite{Witten:1998xy,Gubser:1998fp}
but also branes wrapped around trivial cycles can be stable and BPS.
Such branes carry the same quantum numbers with point-like gravitons,
and are called giant gravitons \cite{McGreevy:2000cw,Grisaru:2000zn,Hashimoto:2000zp,Mikhailov:2000ya}.
The BPS partition function for scalar fields in 4d ${\cal N}=4$ SYM
was reproduced by geometric quantization of
giant gravitons in \cite{Biswas:2006tj},
and the same quantity was also obtained by using the dual giants in \cite{Mandal:2006tk}.
Similar analysis was done in \cite{Bhattacharyya:2007sa} for the 3d and 6d theories realized on M2 and M5-branes.

\begin{figure}[htb]
\centering
\includegraphics{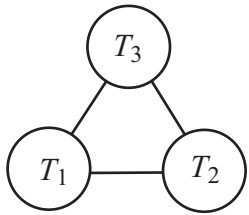}
\caption{A typical structure of the theory realized on a system of giant gravitons.
Each vertex represents the theory $T_i$ with rank $m_i$ realized on
$m_i$ coincident giant gravitons wrapped around cycle $i$.
Edges connecting vertices are degrees of freedom arising on intersections
of cycles.
$d=3$ case is shown.}\label{quiver.eps}
\end{figure}

The contribution of
giant gravitons with different wrapping numbers
to the superconformal index
was suggested by a characteristic form of
the analytic result in \cite{Bourdier:2015wda} 
for the unrefined Schur index of ${\cal N}=4$ SYM.
(See also \cite{Bourdier:2015sga} for generalization to ${\cal N}=2$ quiver gauge theories.)
It has been confirmed
by studying fluctuation modes of giant gravitons
in \cite{Arai:2019xmp,Arai:2020qaj,Imamura:2021ytr} for the ${\cal N}=4$ SYM,
and for more general theories in \cite{Arai:2019wgv,Arai:2019aou,Arai:2020uwd,Fujiwara:2021xgu,Imamura:2021dya}.
In all cases
the index $I_N^{(T)}$ of a theory $T$ with finite $N$
whose holographic dual is string/M theory in $AdS\times M_T$
is given by a multiple expansion of the form
\begin{align}
\frac{I_N^{(T)}}{I_\infty^{(T)}}=
\sum_{m_1=0}^\infty
\cdots
\sum_{m_d=0}^\infty
x_1^{m_1N}\cdots
x_d^{m_dN}
F^{(T)}_{m_1,\ldots,m_d},
\label{multipleexp}
\end{align}
where $m_1,\ldots,m_d$ are wrapping numbers of giant gravitons around
appropriately chosen $d$ supersymmetric cycles in $M_T$.
The number of cycles $d$ depends on the theory $T$.
For each set of the wrapping numbers the function $F^{(T)}_{m_1,\ldots,m_d}$
is the index of the field theory realized on the corresponding system of
giant gravitons and $F_{0,\ldots,0}^{(T)}=1$.
A typical structure of the theory is
shown in Figure \ref{quiver.eps}.
It is the direct product of theories $T_i$ realized on
cycles $i=1,\ldots,d$
coupling to the degrees of freedom
arising on their intersections.
$F^{(T)}_{m_1,\ldots,m_k}$ are $N$-independent, and the $N$-dependence
appears only through the prefactor $x_1^{m_1N}\cdots x_d^{m_dN}$.

Similar expansion of the index is also proposed in \cite{Gaiotto:2021xce} based on
a complementary analysis on the gauge theory side,
and such expansions were named ``the giant graviton expansions''.
See also \cite{Lee:2022vig}.
In their analysis
the correspondents for giant gravitons and fluctuations on them
are baryon operators and their modifications, respectively.
Although there are
some technical subtleties about the relation
between these expansions obtained on the two sides of the AdS/CFT correspondence
they are expected to be essentially the same.

In this work we focus on the contribution of
giant gravitons wrapped around a single cycle $i$.
Let $F^{(T)}_{i,m_i}=F^{(T)}_{0,\ldots,m_i,\ldots,0}$ be such contributions.
It is essentially the index $I_{m_i}^{(T_i)}$ of the theory $T_i$
realized on the worldvolume of $m_i$ coincident giant gravitons.
More precisely, $F_{i,m}^{(T)}$ and $I_m^{(T_i)}$ are
related by a simple variable change of the fugacities
as we will explain shortly.
A purpose of this work is to study this relation in detail.

For concreteness let us consider
the Schur index \cite{Gadde:2011uv} of the ${\cal N}=4$ $U(N)$ SYM.
Let $J_1$ and $J_2$ be the angular momenta and $R_x$, $R_y$, and $R_z$ be
the $R$ charges.
The Schur index is defined by
\begin{align}
I_N^{(D3)}(x,y)=\tr[(-1)^F
q^{J_1}x^{R_x}y^{R_y}],\quad
q=xy,
\label{schurindex}
\end{align}
where we denote the ${\cal N}=4$ $U(N)$ SYM by $T=D3$.
The constraint $q=xy$ is necessary to preserve the supersymmetry.
We eliminate $q$ by the constraint and treat the index as a function of $x$ and $y$.

The giant graviton expansion for the Schur index is given in \cite{Arai:2020qaj}
as the double expansion
\begin{align}
\frac{I_N^{(D3)}}{I^{(D3)}_\infty}=
\sum_{m_x=0}^\infty
\sum_{m_y=0}^\infty
x^{m_xN}y^{m_yN}F^{(D3)}_{m_x,m_y}.
\label{ggexpansion}
\end{align}
The two non-negative integers $m_x$ and $m_y$ are wrapping numbers over
the cycles $i=x$ and $i=y$,
respectively,
which are respectively defined to be the $R_x$ and $R_y$ fixed roci
in $M_{D3}=S^5$.\footnote{We use $x$ and $y$ for the labels of cycles.
These should not be confused with the fugacities.}
A special property of the Schur index is that the functions $F_{m_x,m_y}^{(D3)}$ are factorized \cite{Arai:2020qaj}:
\begin{align}
F_{m_x,m_y}^{(D3)}=(xy)^{m_xm_y}F^{(D3)}_{x,m_x}F^{(D3)}_{y,m_y}.
\label{factorization}
\end{align}
Therefore, we only need to determine $F_{i,m}^{(D3)}$
to write down the expansion (\ref{ggexpansion}).
$F_{i,m}^{(D3)}$ is the index of the ${\cal N}=4$ $U(m)$ SYM realized
on $m$ coincident giant gravitons
wrapped around the cycle $i(=x,y)$.
The theories $T$, $T_x$, and $T_y$ are accidentally the same in this case,
and the corresponding indices $I_N^{(D3)}$, $F_{x,N}^{(D3)}$, and $F_{y,N}^{(D3)}$
are also essentially the same functions.
Careful analysis of the action of superconformal algebra on the $AdS$ boundary
and that on giant gravitons wrapped around the cycle $x$
shows that
they are related by the involution \cite{Arai:2019xmp}
\begin{align}
(H,J_1,J_2,R_x,R_y,R_z)
\rightarrow
(H-2R_x,R_y,R_z,-R_x,J_1,J_2).
\label{involution}
\end{align}
A similar relation also holds for cycle $y$.
This means that $I_N^{(D3)}(x,y)$ and $F_{i,N}^{(D3)}(x,y)$ are
related by the variable changes \cite{Arai:2019xmp}
\begin{align}
F_{x,N}^{(D3)}(x,y)
=I_N^{(D3)}(x^{-1},xy),\quad
F_{y,N}^{(D3)}(x,y)
=I_N^{(D3)}(xy,y^{-1}).
\label{iff}
\end{align}
See also \cite{Gaiotto:2021xce}
for a derivation of (\ref{iff})
on the gauge theory side.

Despite the simplicity of (\ref{iff}),
it is not straightforward to calculate
$F_{i,N}^{(D3)}$ from $I_N^{(D3)}$,
and vice versa,
because we usually do not have analytic form of the
indices, and only series expansions are available.
The variable changes relate different convergence regions of the
functions,
and it is a non-trivial problem to generate one from another.
In this paper, we propose a simple prescription to obtain the functions $F_{i,N}^{(T)}$
from $I_N^{(T_i)}$.

To understand the relation, we need to carefully analyze the dependence of the
series expansion on the choice of expansion variables.
Different choices of expansion variables give different series expansions.
Indeed, the expansion found in \cite{Gaiotto:2021xce}
is the simple expansion over a single non-negative integer $m$:
\begin{align}
\frac{I_N^{(T)}}{I_\infty^{(T)}}=\sum_{m=0}^\infty x^{mN} F_{i,m}^{(T)}.
\label{simpleexp}
\end{align}
This looks different from the multiple expansion (\ref{multipleexp}).
Once we have understood the relation
between $I_N^{(T_i)}$ and $F_{i,N}^{(T)}$,
we can give a partial explanation to this difference.

In the process of confirming the relation (\ref{simpleexp})
the analytic continuation
has been necessarily done in \cite{Gaiotto:2021xce}
for functions appearing in (\ref{simpleexp})
based on a pole structure
commonly found in the functions.
Our prescription is useful to clarify the pole structure
and enable us to study the relations like (\ref{iff}) in a
systematic way.

The paper is organized as follows.

In the next section we prepare tools used in
the following sections.
In particular,
we define ``a domain'' specifying the expansion
variables.
We also define
the plethystic exponential and the plethystic logarithm with emphasis on the
dependence on domains.
In section \ref{cont.sec} we explain the necessity
of the analytic continuation,
and propose
a simple prescription to realize it.
We numerically test the proposed method
in Section \ref{test.sec} for the maximally supersymmetric theories
realized on D3, M5, and M2-branes.
The last section is devoted to
conclusions and discussion.

\section{Plethystic exponential and plethystic logarithm}
\begin{figure}[htb]
\centering
\includegraphics{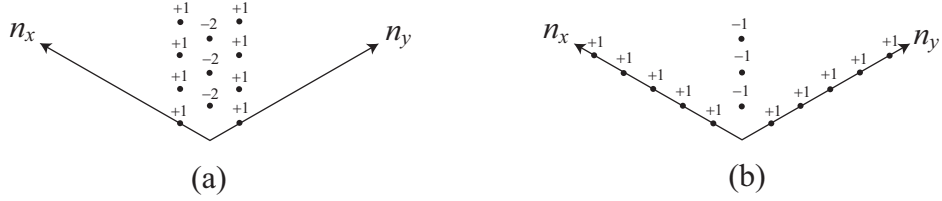}
\caption{(a) The letter index $i_1^{(D3)}$ of the ${\cal N}=4$ vector multiplet.
(b) The letter index $i_\infty^{(D3)}$ of the type IIB supergravity multiplet
in $AdS_5\times S^5$.}\label{ispterms0.eps}
\end{figure}

The calculation of the index $I$
of a free field theory usually starts from
the analysis of single-particle states.
The corresponding index is called the letter index.
It is obtained by summing up the contributions from all one-particle states.
If there are two fugacities $x$ and $y$
it is given by\footnote{We mainly consider the case with two variables.
Generalization is straightforward.}
\begin{align}
i=\sum_{(n_x,n_y)\in R}c_{n_x,n_y}x^{n_x}y^{n_y},
\label{mnexp}
\end{align}
where $R$ is a certain region on the $2$-dimensional charge lattice
specifying a set of monomials appearing
in the expansion.
We call such a region ``a domain.''
We usually adopt $R$ such that
it covers all one-particle states.
We call such a domain ``a physical domain''.
For the Schur index of $T=D3$ we can take
\begin{align}
R_0=\{(n_x,n_y)|n_x,n_y\geq0\}\backslash(0,0)
\end{align}
as a physical domain.
Later we will discuss other choices of $R$.
We call the series associated with a domain $R$ the $R$-series.
For a free field theory the series sums up into a simple rational function.
For example, the letter index of the ${\cal N}=4$ $U(1)$ vector multiplet is
\cite{Kinney:2005ej}
\begin{align}
i^{(D3)}_1=1-\frac{(1-x)(1-y)}{1-xy},
\label{ispvec}
\end{align}
and
that of the type IIB supergravity multiplet in $AdS_5\times S^5$,
which is dual to the large $N$ SYM, is
\cite{Kinney:2005ej}
\begin{align}
i^{(D3)}_\infty=
\frac{1}{1-x}
+\frac{1}{1-y}
-\frac{1}{1-xy}
-1.
\end{align}
By expanding these functions
into terms appearing in the domain $R_0$
we can obtain the set of coefficients
in (\ref{mnexp}).
In the following we use the notation $i|_R$ when we want to emphasize
that $i$ is given as an $R$-series.
We graphically express a series by plotting the coefficients
in the two-dimensional lattice
as shown in Figure \ref{ispterms0.eps}.

Once we have obtained the letter index (\ref{mnexp}),
the index for multi-particle states is uniquely determined
by considering the combinatorics of the letters.
The solution to this problem is
\begin{align}
I=\Pexp i,
\label{iandi}
\end{align}
where $\Pexp$ is the plethystic exponential defined by
\begin{align}
\Pexp f(x,y)=\exp\left(\sum_{m=1}^\infty\frac{1}{m}f(x^m,y^m)\right).
\label{defpexp}
\end{align}
For the definition (\ref{defpexp}) to make sense
the condition $|x^{n_x}y^{n_y}|<1$ must hold
for all terms appearing
in the sum (\ref{mnexp}).
This means all terms in $f$
must be contained in the domain $R_{(x,y)}$ defined by
\begin{align}
R_{(x,y)}:=\{(n_x,n_y)|n_x\log|x|+n_y\log|y|<0\}.
\label{halfregion}
\end{align}
For example, if we take $x$ and $y$ satisfying $|x|=|y|<1$,
(\ref{halfregion}) gives the upper half plane with the horizontal boundary excluded,
and all points of $i_1^{(D3)}$ and $i_\infty^{(D3)}$ shown in Figure \ref{ispterms0.eps} are contained in the domain.
(Namely, we can use $R_{(x,y)}$ as a physical domain.)
If $i$ is expanded with such a domain $R$,
$I$ given by (\ref{iandi})
can also be expanded in a similar form;
\begin{align}
I=1+\sum_{(n_x,n_y)\in R}c'_{n_x,n_y}x^{n_x}y^{n_y}
\label{pexpexp}
\end{align}
with the same domain $R$.
We can show that
the map from $\{c_{n_x,n_y}\}$ to $\{c'_{n_x,n_y}\}$
is one to one if
\begin{itemize}
\item $R$ is a sector with the center angle $\theta\leq\pi$.
\item $R\cap(-R)=\phi$.
\end{itemize}
The latter is the condition for the boundary of $R$.
We do not include the origin, and if $\theta=\pi$ we
can include at most one of
two boundary rays.
We always require $R$ to satisfy
these conditions,
and then we can define the inverse operation to $\Pexp$:
\begin{align}
i=\Plog I.
\end{align}
This is called the plethystic logarithm.
It is important that we can introduce the group
structure for $\Pexp i$ under
the multiplication,
which corresponds to the addition for $i$.

Even for an interacting theory we
can calculate the index by using localization method
if it is a Lagrangian theory.
For the ${\cal N}=4$ SYM with $U(N)$ gauge group
the index is given by
\begin{align}
I^{(D3)}_N|_{R_{(x,y)}}
=\frac{1}{N!}\int\prod_{a=1}^N\frac{dz_a}{2\pi iz_a}\prod_{a\neq b}\left(1-\frac{z_a}{z_b}\right)
\Pexp\left(i_1^{(D3)}|_{R_{(x,y)}} \sum_{a,b}\frac{z_a}{z_b}\right),
\label{localization}
\end{align}
where $z_a$ ($a=1,\ldots,N$) are gauge fugacities.
As is explicitly shown
we suppose $i_1^{(D3)}$ is given as an $R_{(x,y)}$-series.
Usually $R_{(x,y)}$ is assumed to be a physical domain.
This means that
$|x^{n_x}y^{n_y}|<1$ is satisfied for all terms in $i_1^{(D3)}|_{R_{(x,y)}}$.
If we take the unit circle $|z_a|=1$ for the integration contours
$|x^{n_x}y^{n_y}\frac{z_a}{z_b}|<1$ is also satisfied,
and $\Pexp$ in
(\ref{localization}) makes sense.
The contour integrals extract the contribution
of gauge invariant states
and give $I_N$ as an $R_{(x,y)}$-series.

\section{Variable changes and analytic continuation}\label{cont.sec}
In the following we frequently use the variable changes in (\ref{iff}),
and it is convenient to introduce $\sigma_x$ and $\sigma_y$ to represent these
variable changes as follows.
\begin{align}
\sigma_x(x,y,q)=(x^{-1},q,y),\quad
\sigma_y(x,y,q)=(q,y^{-1},x).
\label{varchangex}
\end{align}
If we use the triangular lattice to express the
series expansion as in Figure \ref{ispterms0.eps},
$\sigma_x$ and $\sigma_y$ are the reflections through the lines
perpendicular to $n_x$ and $n_y$ axes, respectively.
With the maps $\sigma_i$ ($i=x,y$) the relations in (\ref{iff}) are expressed in the simple form
\begin{align}
F_{i,N}^{(D3)}=\sigma_iI_N^{(D3)}\quad(i=x,y).
\label{fn0f0n}
\end{align}
We want to confirm that $F_{i,N}^{(D3)}$ given by
(\ref{fn0f0n}) correctly reproduce $I_N^{(D3)}$ via (\ref{ggexpansion}).
Let $R$ be a physical domain used for the expansion of $I_N^{(D3)}$.
To confirm the relation (\ref{ggexpansion}) holds
we need $F_{i,N}^{(D3)}|_R$.
However, the variable change $\sigma_i$ transforms $I_N^{(D3)}|_R$
into $F_{i,N}^{(D3)}|_{\sigma_iR}$, where
$\sigma_iR$ is the image of $R$ under the map $\sigma_i$.
Therefore,
to compare the left hand side and the right hand side of (\ref{ggexpansion}),
we need to resum or analytically continue
$F_{i,N}^{(D3)}|_{\sigma_iR}$
to $F_{i,N}^{(D3)}|_{R}$.

In general,
if we have $i|_R$ for a domain $R$,
it can also be regarded as $i|_{R'}$ for
another domain $R'$ which contains $R$ as a subset.
Therefore,
it is convenient to define ``a maximal domain''
which cannot be enlarged any more.
$R$ is a maximal domain if
\begin{align}
R\cup(-R)\cup O=L,
\end{align}
where $O$ is the origin and $L$ is the whole plane.
A maximal domain $R$ is a sector with center angle $\pi$,
and only one of the boundary rays is included in $R$.

Let us suppose that $f|_{R'}$ for
a maximal domain $R'$ is given
and we want to obtain $(\Pexp f)|_R$ for another
maximal domain $R\neq R'$.
We divide $R'$ into the following two parts.
\begin{align}
R_1=R'\backslash R,\quad
R_2=R'\cap R.
\end{align}
By definition, $R'=R_1\cup R_2$ and $R=(-R_1)\cup R_2$.
Correspondingly, we divide $f$ into two parts $f_1$ and $f_2$
so that $f|_{R'}=f_1|_{R_1}+f_2|_{R_2}$.
The plethystic exponential of $f|_{R'}$ is factorized into two
factors:
\begin{align}
\Pexp f|_{R'}=(\Pexp f_1|_{R_1})(\Pexp f_2|_{R_2}).
\label{pexpfactorization}
\end{align}
The factor $\Pexp f_2|_{R_2}$ can be regarded
as an $R$-series.
What we need to do to obtain $(\Pexp f)|_R$ is
to rewrite the other factor
$\Pexp f_1|_{R_1}$ as a $(-R_1)$-series.
This can be done by using analytic continuation as follows.
Let us suppose that $f_1$ is given by
\begin{align}
f_1(x,y)=\sum_{(n_x,n_y)\in R_1}c_{n_x,n_y}x^{n_x}y^{n_y}.
\end{align}
For values of $x$ and $y$ such that
$R_1\subset R_{(x,y)}$
we can rewrite the plethystic exponential in the
product form
\begin{align}
\Pexp f_1|_{R_1}=\prod_{(n_x,n_y)\in R_1}\frac{1}{(1-x^{n_x}y^{n_y})^{c_{n_x,n_y}}}.
\label{pexpa}
\end{align}
Once we have obtained this expression, we can analytically continue
this function to values of $x$ and $y$ such that
$-R_1\subset R_{(x,y)}$.
Then, we can rewrite
(\ref{pexpa}) as
\begin{align}
\prod_{(n_x,n_y)\in R_1}\frac{(-x^{n_x}y^{n_y})^{-c_{n_x,n_y}}
}{(1-x^{-n_x}y^{-n_y})^{c_{n_x,n_y}}}
=P(x,y)\Pexp f_1(x^{-1},y^{-1})|_{-R_1},
\label{contprod}
\end{align}
where $P(x,y)$ is the monomial function
\begin{align}
P(x,y)=\prod_{(n_x,n_y)\in R_1}(-x^{n_x}y^{n_y})^{-c_{n_x,n_y}}.
\label{prefactor}
\end{align}
It is important that if $f_1$
has infinitely many terms
(\ref{prefactor}) may not be well-defined,
and then our prescription does not work.

By combining
(\ref{contprod})
with the factor $\Pexp f_2$ we obtain
the analytically continued plethystic exponential
associated with the domain $R$:
\begin{align}
\Pexp_R f\equiv P\Pexp f^{\rm mod}|_R
\end{align}
where $f^{\rm mod}$ is defined by
\begin{align}
f^{\rm mod}(x,y)=f_1(x^{-1},y^{-1})+f_2(x,y).
\end{align}
Each point $(n_x,n_y)\in R_1$ in $f$ is moved to the opposite point $(-n_x,-n_y)\in -R_1$
in the modified function $f^{\rm mod}$,
and all terms of $f^{\rm mod}$ are contained in $R$.

In fact, essentially the same prescription
has been used in previous works.
Because giant gravitons are wrapped around
topologically trivial cycles,
there are unwrapping modes with negative excitation energies.
Such modes are handled
in \cite{Arai:2019xmp,Arai:2020qaj,Imamura:2021ytr}
by applying the prescription explained above
to the integrand of the gauge fugacity integrals.
The variable change of the index after the gauge fugacity integrals
has also been considered in \cite{Gaiotto:2021xce}.
Although detailed explanation is not given,
the pole structure caused by the factor
(\ref{pexpa}) was pointed out,
and essentially the same method
seems to be used.

The prescription explained above can be summarized in the following equations.
\begin{align}
F_{i,N}^{(T)}|_R=\Pexp_R\sigma_ii_N^{(T_i)},\quad
i_N^{(T_i)}=\Plog I_N^{(T_i)}.
\label{proposal}
\end{align}
We first calculate $i_N^{(T)}$ as the plethystic logarithm
of $I_N^{(T_i)}$,
and then we calculate the
analytically continued plethystic exponential.

Unfortunately, it will turn out that this prescription works only in limited cases.
If $\sigma_ii_N^{(T_i)}$ has infinitely many points outside the domain $R$,
then the factor
(\ref{prefactor})
becomes an infinite product which in general does not converge.
Even so, we will find that (\ref{proposal})
is quite effective for simple expansion (\ref{simpleexp}).

\section{Numerical tests}\label{test.sec}
In this section we numerically calculate $F_{i,N}^{(T)}$ from $I_N^{(T_i)}$
by (\ref{proposal}) in some examples
and confirm the consistency with known results.
The main results in \ref{simpleschur} and \ref{simplesci}
have been already given in \cite{Gaiotto:2021xce}.

\subsection{Double expansion of the Schur index for $T=D3$}
\begin{figure}[htb]
\centering
\includegraphics{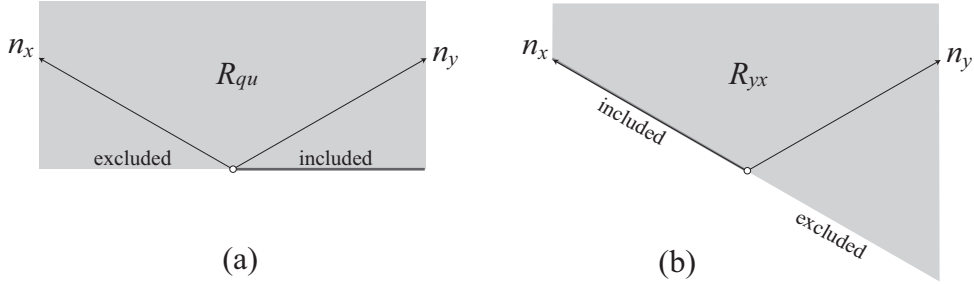}
\caption{The maximal domains $R_{qu}$ and $R_{yx}$}\label{hqu.eps}
\end{figure}

We first consider the double giant graviton expansion
of the Schur index of $T=D3$ studied in \cite{Arai:2020qaj}.
Let us define $\fq$ and $u$ by
\begin{align}
x=\fq u,\quad
y=\fq u^{-1},\quad
q=\fq^2.
\end{align}
In \cite{Arai:2020qaj} functions $F_{i,N}^{(D3)}$ are treated as $\fq$-series and the coefficients are given as rational functions
of $u$.\footnote{The fugacity $\fq$ is denoted by $q$ in \cite{Arai:2020qaj}.}
Here we do the further $u^{-1}$ expansion ($u$ expansion around $u=\infty$) after the $\fq$-expansion.
This corresponds to the maximal physical domain
(Figure \ref{hqu.eps} (a))
\begin{align}
R_{qu}
&=\lim_{\fq\rightarrow 0,|u|>1}R_{(x,y)}
\nonumber\\
&=\{(n_x,n_y)|n_x+n_y>0\}\cup
\{(n_x,n_y)|n_x+n_y=0,n_y>0\}.
\label{rqudef}
\end{align}

\begin{figure}[htb]
\centering
\includegraphics{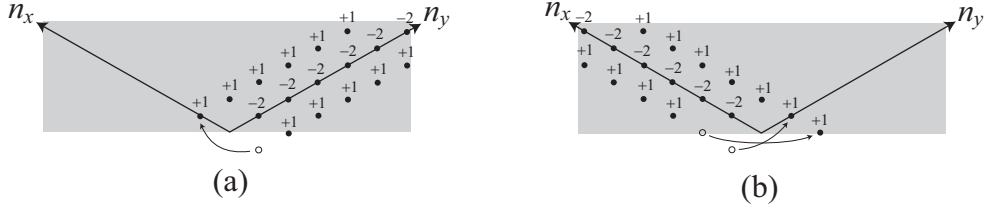}
\caption{The modification of letter indices $\sigma_xi_1^{(D3)}$ and $\sigma_yi_1^{(D3)}$}\label{qu2.eps}
\end{figure}

\begin{figure}[htb]
\centering
\includegraphics{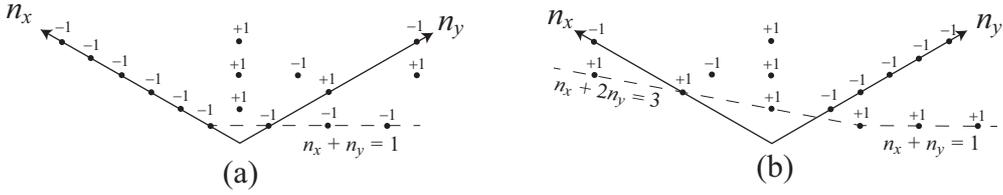}
\caption{(a) $F_{x,1}^{(D3)}|_{R_{qu}}$ (b) $F_{y,1}^{(D3)}|_{R_{qu}}$}\label{gg1.eps}
\end{figure}

Let us first consider $F_{x,1}^{(D3)}=\Pexp(\sigma_xi_1^{(D3)})$ and $F_{y,1}^{(D3)}=\Pexp(\sigma_yi_1^{(D3)})$.
We find $\sigma_xi_1^{(D3)}$ has one term $x^{-1}=\fq^{-1}u^{-1}$ that is not contained in $R_{qu}$, and
we have to move it to the opposite point to define
the modified letter index.
For $\sigma_yi_1^{(D3)}$ we need to move two points corresponding to $xy^{-1}=u^2$ and $y^{-1}=\fq^{-1}u$.
See Figure \ref{qu2.eps}.
Following the prescription explained in the previous section,
we can calculate $F_{x,1}^{(D3)}|_{R_{qu}}$ and $F_{y,1}^{(D3)}|_{R_{qu}}$.
The results are shown in Figure \ref{gg1.eps}.
These agree with the results in \cite{Arai:2020qaj}:\footnote{$F_{x,N}(x,y)$ is denoted by $F_N(q,u)$ in $\cite{Arai:2020qaj}$
and $F_{y,N}(x,y)$ corresponds to $F_N(q,u^{-1})$.
$F_N$ are given in $\cite{Arai:2020qaj}$ as
$\fq$-series.
We need further $u^{-1}$ expansion to obtain (\ref{f10}),
(\ref{f2}),
(\ref{fy2}), and
(\ref{f34}).}
\begin{align}
F_{x,1}^{(D3)}
&=(-u-u^{-1}-u^{-3}-\cdots)\fq
\nonumber\\
&+(-u^2+1)\fq^2+(-u^3+u^{-3})\fq^3
+(-u^4+1-u^{-2}+u^{-6})\fq^4+\cdots
,\nonumber\\
F^{(D3)}_{y,1}&=(u^{-3}+u^{-5}+u^{-7}+\cdots)\fq
\nonumber\\
&+(1-u^{-2})\fq^2+(u^3-u^{-3})\fq^3
+(u^6-u^2+1-u^{-4})\fq^4+\cdots.
\label{f10}
\end{align}
This is not surprising
because essentially the same prescription
is used in \cite{Arai:2020qaj} for the single giant graviton sector.

\begin{figure}[htb]
\centering
\includegraphics{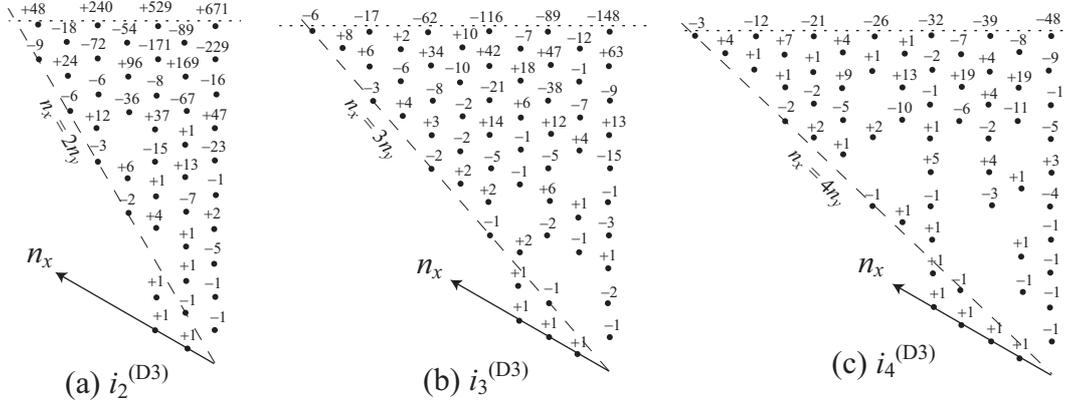}
\caption{$i_N^{(D3)}$ for $N=2,3,4$ are shown.
The distributions are left-right symmetric, and
we show only the left half ($n_x\geq n_y$) of each plot.
The dotted lines show a numerical cut-off $n_x+n_y\leq20$.}\label{n4schurplog.eps}
\end{figure}

Next, let us apply
(\ref{proposal}) to $F_{i,N}^{(D3)}$ with $N\geq 2$.
We first calculate $I_N^{(D3)}$ by (\ref{localization}),
and calculate their plethystic logarithms $i_N^{(D3)}=\Plog I_N^{(D3)}$.
$i_N^{(D3)}$ have information not only about letters
generating the operator spectra but also about non-trivial
constraints among letters called ``syzygies'' \cite{Benvenuti:2006qr}.
The results for $N=2,3,4$ are shown in Figure \ref{n4schurplog.eps}.
The asymptotic behavior of the distribution of dots is important
for the following analysis.
We find in the plots in Figure \ref{n4schurplog.eps}
that except for finite number of dots near the origin
almost all dots are contained in the sectors
bounded by $E_1:n_y= Nn_x$ and $E_2:n_x= Nn_y$.

\begin{figure}[htb]
\centering
\includegraphics{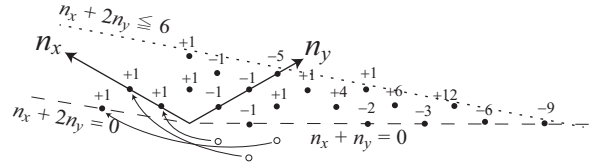}
\caption{The modification of $\sigma_xi_2^{(D3)}$ for $R_{qu}$.
The dotted line shows a numerical cut-off.}\label{sxi2d3.eps}
\end{figure}

\begin{figure}[htb]
\centering
\includegraphics{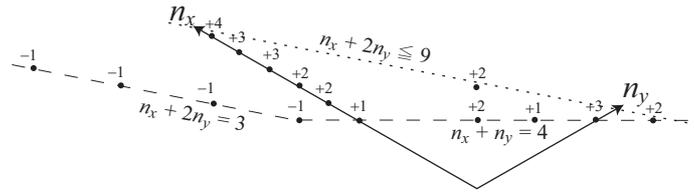}
\caption{$F_{x,2}^{(D3)}|_{R_{qu}}$ is shown.
The dotted line shows a numerical cut-off.}\label{aipexp.eps}
\end{figure}

For $N=2$
the image of $E_2$ under $\sigma_x$, $\sigma_xE_2$, coincides
with the right half of the horizontal line, which is the boundary of $R_{qu}$
(Figure \ref{sxi2d3.eps}).
There are three points outside $R_{qu}$
corresponding to terms $x^{-1}=\fq^{-1}u$, $x^{-2}=\fq^{-2}u^2$,
and $x^{-2}y=\fq^{-1}u^3$.
These give the factor $P=-\fq^4u^{-6}=-x^5y^{-1}$,
and the analytically continued plethystic exponential is shown in Figure \ref{aipexp.eps}.
Again, we find agreement with the corresponding result in \cite{Arai:2020qaj}:
\begin{align}
F_{x,2}^{(D3)}&=
(-u^6+u^4+2+u^{-2}+3u^{-4}+2u^{-6}+\cdots)\fq^4
\nonumber\\
&+(-u^9+2u^5)\fq^5
+(-u^{12}+2u^6+2)\fq^6+\cdots.
\label{f2}
\end{align}

Next, let us consider $F_{y,2}^{(D3)}$.
This time,
$\sigma_yE_1$ is the semi-infinite horizontal line,
and unlike the previous case
the line is not contained in $R_{qu}$.
This means that there are infinitely many points outside $R_{qu}$,
and the factor (\ref{prefactor}) becomes infinite product.
Therefore, (\ref{proposal}) does not work.
In fact, this is also expected from $F_{y,2}^{(D3)}$ in \cite{Arai:2020qaj}:
\begin{align}
F_{y,2}^{(D3)}(\fq,u)&=(2u^{-10}+u^{-12}+3u^{-14}+\cdots)\fq^4+\cdots.
\label{fy2}
\end{align}
As we see, the leading term has coefficient $2$,
and it is impossible to obtain such a series
by the plethystic exponential.

For $N=3$ and $4$ the situation is worse.
Both $\sigma_x$ and $\sigma_y$ maps
one of $E_1$ and $E_2$
below the boundary of $R_{qu}$,
and there are infinitely many points outside $R_{qu}$,
which make the factor (\ref{prefactor}) ill-defined.
We can also see that the leading term coefficients of the
functions in \cite{Arai:2020qaj}
are not $\pm1$.
\begin{align}
F_{x,3}^{(D3)}&=(-2u^{15}+u^{13}+\cdots)\fq^9+\cdots,\nonumber\\
F_{y,3}^{(D3)}&=(5u^{-21}+2u^{-23}+\cdots)\fq^9+\cdots,\nonumber\\
F_{x,4}^{(D3)}&=(-5u^{28}+2u^{26}+\cdots)\fq^{16}+\cdots,\nonumber\\
F_{y,4}^{(D3)}&=(14u^{-36}+5u^{-38}+\cdots)\fq^{16}+\cdots.
\label{f34}
\end{align}
These expansions cannot be reproduced
by the relation (\ref{proposal}).

\subsection{Simple expansion of the Schur index for $T=D3$}\label{simpleschur}
Let us consider the simple giant graviton expansion investigated in \cite{Gaiotto:2021xce}.
The index is treated in \cite{Gaiotto:2021xce} as a series obtained by
the successive expansions in which $y$ expansion is carried out first
and then $x$ expansion is done for the coefficients of the first expansion.
The associated maximal physical domain is (Figure \ref{hqu.eps} (b))
\begin{align}
R_{yx}=\lim_{y\rightarrow 0,|x|<1}R_{(x,y)}
=\{(n_x,n_y)|n_y>0\}\cup
\{(n_x,0)|n_x>0\}.
\label{hyx}
\end{align}

Let us start with the analysis of $F_{x,1}^{(D3)}|_{R_{yx}}$ and $F_{y,1}^{(D3)}|_{R_{yx}}$.

\begin{figure}[htb]
\centering
\includegraphics{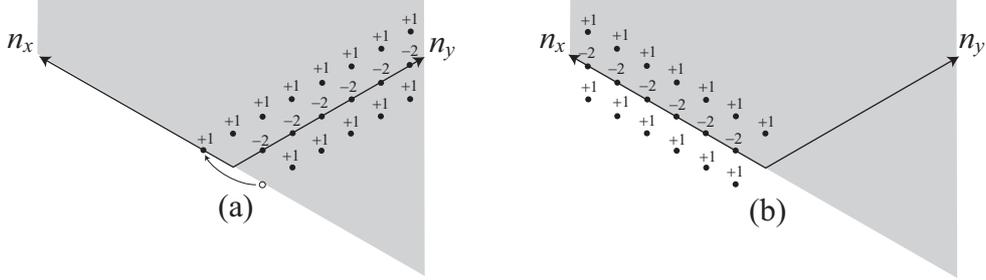}
\caption{(a) The modification of $\sigma_xi_1$. (b) $\sigma_yi_1$}\label{glexp.eps}
\end{figure}
For $\sigma_xi_1^{(D3)}$, there is one dot outside $R_{yx}$ (Figure \ref{glexp.eps} (a)).
This is the same as in Figure \ref{qu2.eps} (a), and hence we obtain the same result
for $F_{x,1}^{(D3)}$ as the previous subsection (Figure \ref{gg1.eps} (a)).
However, for $\sigma_yi_1^{(D3)}$, there are infinitely many points outside $R_{yx}$
corresponding to terms $y^{-1}x^k$ ($k=0,1,2,\ldots$) (Figure \ref{glexp.eps} (b)),
and (\ref{prefactor}) becomes
\begin{align}
P=\prod_{k=0}^\infty(-yx^{-k}).
\end{align}
Although this is an infinite product,
we can give it significance.
$P$ includes the factor $y^{+\infty}$,
and the contribution decouples.
We can simply treat $F_{y,1}^{(D3)}|_{R_{yx}}$ as $0$.
Namely, as far as single-wrapping giant graviton contributions are concerned,
the difference between the multiple expansion (\ref{multipleexp})
and the simple expansion (\ref{simpleexp})
comes from the different choice of the domains.

Let us proceed to $N\geq 2$ contributions.
For $\sigma_yi_N^{(D3)}$, we find infinitely many points outside $R_{qu}$,
just like the $N=1$ case.
Because both positive and negative coefficients appear,
it is not clear whether they decouple like $F_{y,1}^{(D3)}$.
Here, based on the analysis in \cite{Gaiotto:2021xce}, we simply assume
their decoupling and focus only on $F_{x,N}^{(D3)}$.

\begin{figure}[htb]
\centering
\includegraphics{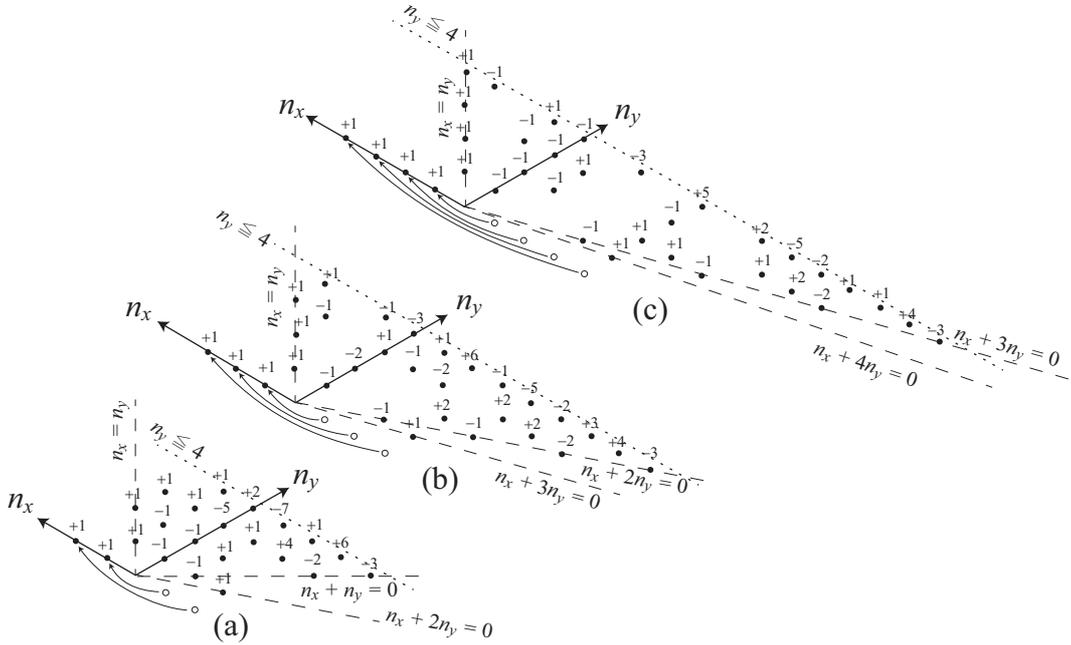}
\caption{The modification of $i_N^{(D3)}$ for $R_{yx}$ for $N=2,3,4$.
(a) $\sigma_xi_2$. (b) $\sigma_xi_3$. (c) $\sigma_xi_4$.
The dotted lines show a numerical cut-off}\label{sxi234.eps}
\end{figure}

\begin{figure}[htb]
\centering
\includegraphics{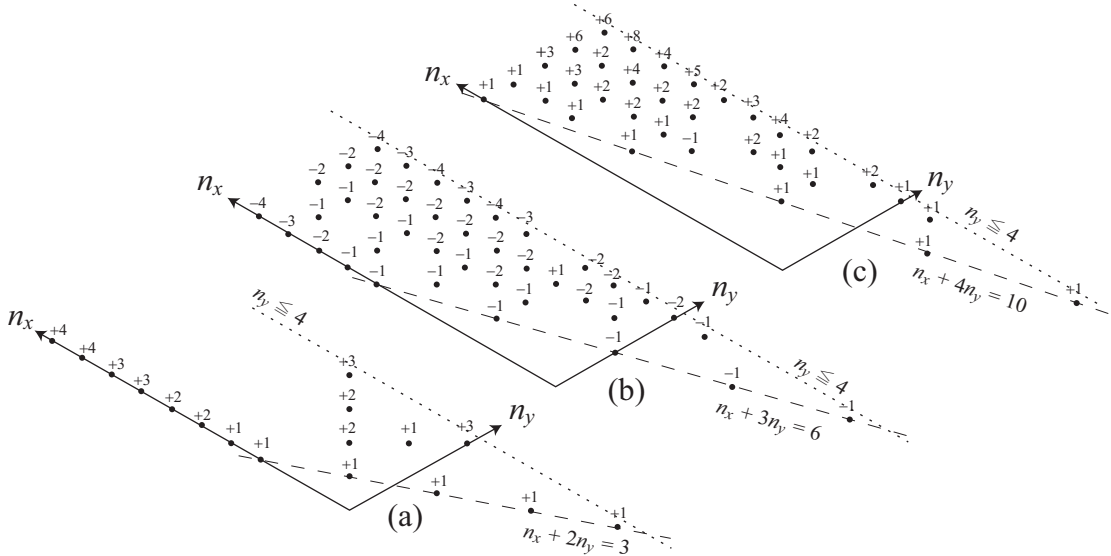}
\caption{
(a) $F_{x,2}^{(D3)}|_{R_{yx}}$.
(b) $F_{x,3}^{(D3)}|_{R_{yx}}$.
(c) $F_{x,4}^{(D3)}|_{R_{yx}}$.
The dotted lines show a numerical cut-off.
}\label{gg234.eps}
\end{figure}

\begin{figure}[htb]
\centering
\includegraphics{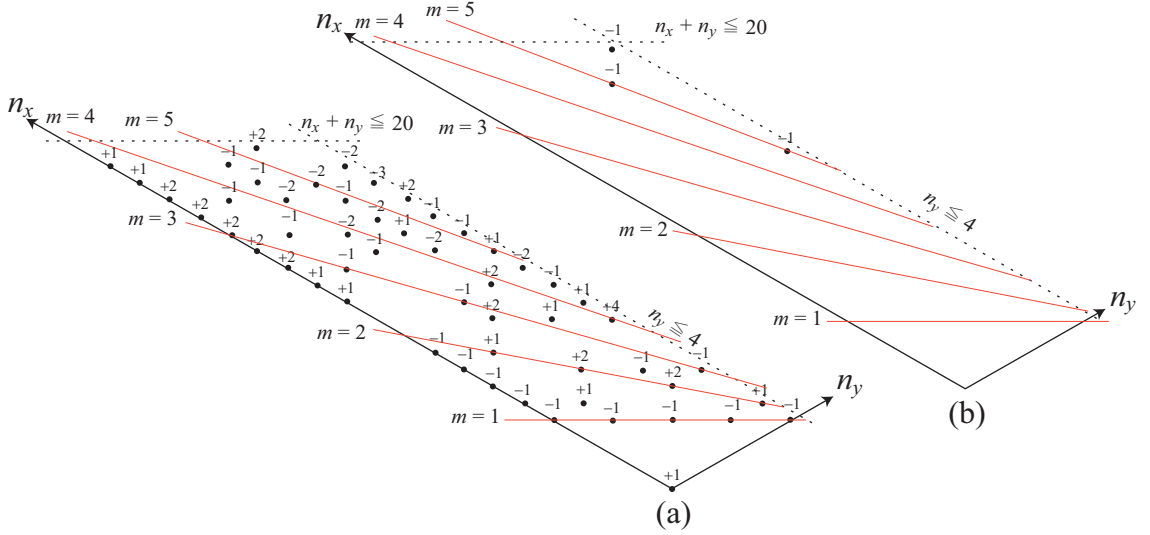}
\caption{
The result of numerical check for $N=3$.
(a) The ratio $I_3^{(D3)}/I_\infty^{(D3)}$.
(b) $(I_3^{(D3)}/I_\infty^{(D3)})-\sum_{m=0}^4x^{mN}F_{x,m}^{(D3)}$.
The dotted lines show numerical cut-offs.
The red lines labeled by $m$ show thresholds for respective wrapping numbers.
Up to expected errors due to the $m=5$ contribution
indicated by the line labeled by $m=5$, all terms are correctly canceled.
}\label{ggtest.eps}
\end{figure}

In $i_N^{(D3)}$ shown in Figure \ref{n4schurplog.eps}
we find $N$ dots on the positive part of the $n_x$ axis
corresponding to the $1/2$ BPS operators $\tr X^k$ ($k=1,2,\ldots,N$).
By $\sigma_x$ they are mapped to the $N$ points
on the negative part of the $n_x$ axis.
Because they are not contained in $R_{yx}$
we need to move them back to the opposite points
on the positive part of the $n_x$ axis (Figure \ref{sxi234.eps}).
The factorization corresponding to (\ref{pexpfactorization}) is
\begin{align}
I_N^{(D3)}(x,y)=\frac{1}{\prod_{k=1}^N(1-x^k)}I'^{(D3)}_N(x,y),
\end{align}
The fractional factor coming from the $N$ points on the $n_x$ axis
and produces
poles on the $x$-plane at $k$-roots of unity with $1\leq k\leq N$.
As is pointed out in \cite{Gaiotto:2021xce} these are only
poles we need to take care of in the analytic continuation.
The function $I'^{(D3)}_N$ corresponds to
all other points with $n_y\geq1$
in the range $0\leq n_x\leq(N+1)n_y$.
The coefficient of $y$-expansion $I'^{(D3)}_N$ at order $n_y$
is a Laurant polynomial of $x$ consisting
of terms in this range.
Therefore, we can safely perform the variable change $\sigma_x$ for the function
$I'^{(D3)}_N$.
After the variable change we obtain
\begin{align}
F_{x,N}^{(D3)}=\frac{(-1)^Nx^{\frac{N(N+1)}{2}}}{\prod_{k=1}^N(1-x^k)}\sigma_xI'^{(D3)}_N
\end{align}
where
$\sigma_xI'^{(D3)}_N$ has terms in the range $-Nn_y\leq n_x\leq n_y$ for each order of its $y$-expansion.
Including the fractional factor with the numerator $(-1)^Nx^{\frac{N(N+1)}{2}}$,
we obtain $F_{x,N}^{(D3)}|_{R_{yx}}$ consisting of terms in the region (Figure \ref{gg234.eps})
\begin{align}
\frac{N(N+1)}{2}-Nn_y
\leq n_x.
\end{align}
As is confirmed in \cite{Gaiotto:2021xce},
the simple expansion (\ref{simpleexp})
correctly reproduces the index $I_N^{(D3)}$ for different values of $N$.
See Figure \ref{ggtest.eps}.

\subsection{Superconformal index for $T=D3$}\label{simplesci}
In the analysis of the Schur index
we could focus on $F_{i,N}^{(D3)}|_R$
thanks to the factorization
(\ref{factorization}).
This is not the case
for the superconformal index without the Schur limit taken.
The superconformal index
of the ${\cal N}=4$ SYM is defined by \cite{Kinney:2005ej} 
\begin{align}
I^{(D3)}=\tr[(-1)^F
q^{J_1}p^{J_2}x^{R_x}y^{R_y}z^{R_z}],\quad
qp=xyz.
\label{scidef}
\end{align}
The giant graviton expansion investigated in \cite{Imamura:2021ytr} is
a triple expansion with functions
$F_{m_x,m_y,m_z}^{(D3)}$ in the summand,
and the involution (\ref{involution})
gives the following
relation between $F_{x,N}^{(D3)}=F_{N,0,0}^{(D3)}$ and $I_N^{(D3)}$ \cite{Arai:2019xmp,Gaiotto:2021xce}:
\begin{align}
F_{x,N}^{(D3)}(p,q,x,y,z)=I_N^{(D3)}(y,z,x^{-1},p,q).
\end{align}
Let $\sigma_x$ denote this variable change.
$F_{0,N,0}^{(D3)}$ and $F_{0,0,N}^{(D3)}$ are also related to $I_N^{(D3a)}$ via
similar variable changes, which we denote by $\sigma_y$ and $\sigma_z$, respectively.

Due to the absence of the factorization,
we need to calculate functions $F_{m_x,m_y,m_z}^{(D3)}$ directly
by using localization formula
similar to (\ref{localization}).
The letter index appearing in the integrand of the localization formula
contains
$\sigma_xi_1^{(D3)}$, $\sigma_yi_1^{(D3)}$, and $\sigma_zi_1^{(D3)}$ at the same time,
where $i_1^{(D3)}$ is given by
\cite{Kinney:2005ej}
\begin{align}
i_1^{(D3)}=1-\frac{(1-x)(1-y)(1-z)}{(1-q)(1-p)}.
\end{align}
Even if we use a maximal domain we cannot cover three functions $\sigma_ii_1^{(D3)}$
with a single domain, and we need to use deformed contours
for the gauge fugacity integrals \cite{Imamura:2021ytr,Lee:2022vig}.

\begin{figure}[htb]
\centering
\includegraphics{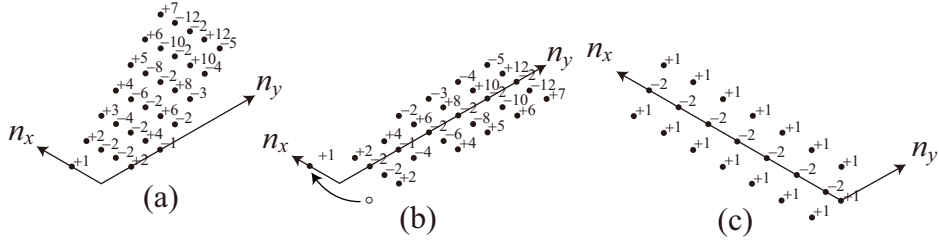}
\caption{(a) The letter index $i_1^{(D3)}$
(b) The modification of $\sigma_xi_1^{(D3)}$ for $R_{yx}$.
(c) $\sigma_yi_2^{(D3)}$ and $\sigma_zi_2^{(D3)}$ are identical after the
projection to the two-dimensional lattice.
}\label{sciletter.eps}
\end{figure}

\begin{figure}[htb]
\centering
\includegraphics{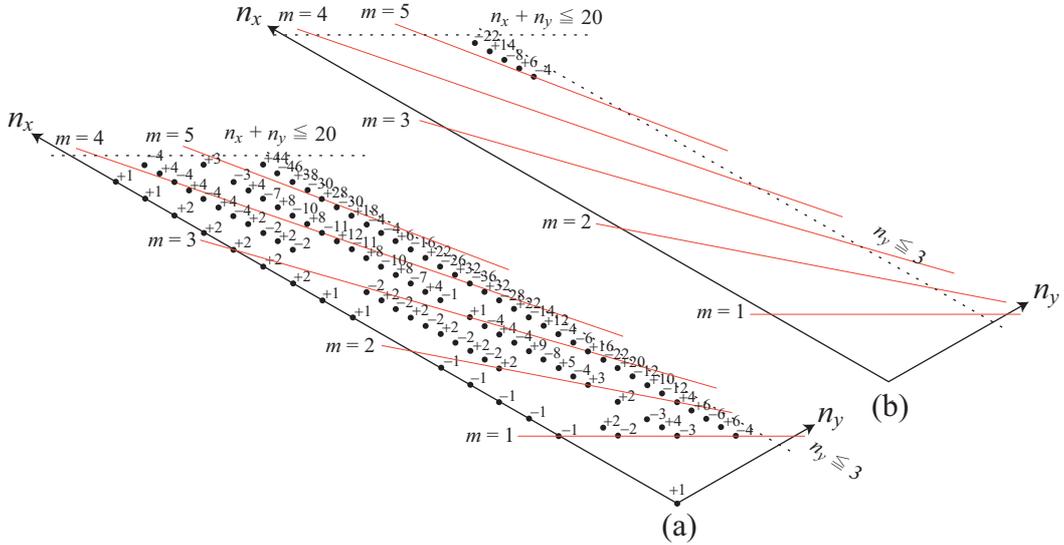}
\caption{(a) The ratio of the superconformal indices $I_N^{(D3)}/I_\infty^{(D3)}$ for $N=3$.
(b) $(I_N^{(D3)}/I_\infty^{(D3)})-\sum_{m=0}^4x^{mN}F_{x,N}^{(D3)}$ for $N=3$.
The dotted lines show numerical cut-offs.
The red lines labeled by $m$ show thresholds for respective wrapping numbers.
Only small number of dots, which are consistent with the higher order contributions
with $m\geq5$, are left.
}\label{scitest.eps}
\end{figure}

Despite the absence of the factorization,
the analysis in \cite{Gaiotto:2021xce} suggests that
only $F_{x,m}^{(D3)}$ contribute to the index if we use an appropriate
domain, and then our relation (\ref{proposal}) is
still useful to obtain
$F_{x,N}^{(D3)}$ from $I_N^{(D3)}$.
Corresponding to four independent variables we need to use
a four-dimensional lattice to express full expansion of the
index.
To make two-dimensional plot possible
we project the four-dimensional lattice onto the two-dimensional lattice
corresponding to the unrefinement $y=z$ and $p=q$.
Then, the index becomes function of $x$ and $y$
just like the Schur index.\footnote{The projection is
used only when we show functions as two-dimensional plots,
and we do not take any unrefinement in the calculation.
We regard the index as a function of $x$, $\sqrt{yz}$, $y/z$, and $q/p$,
and treat $\sqrt{yz}$ as if $y$ in the previous subsection.
Although we neglect $y/z$ and $q/p$ in the graphical expression,
we keep all variables in the calculation.
}
The letter indices $i_1^{(D3)}$ and $\sigma_ii_1^{(D3)}$ ($i=x,y,z$)
after the projection
are shown in Figure \ref{sciletter.eps}.
We see that if we use the domain $R_{yx}$
$F_{y,1}^{(D3)}$ and $F_{z,1}^{(D3)}$ decouple
 just like the Schur index,
and $F_{x,1}^{(D3)}$ is the only contribution with $m_x+m_y+m_z=1$.
Although we cannot prove the decoupling of all $F_{m_x,m_y,m_z}^{(D3)}$ with
$m_y+m_z\geq1$, the analysis in \cite{Gaiotto:2021xce} shows that
we can reproduce the finite $N$ index
by the simple expansion containing only $F_{x,m}^{(D3)}$.
See Figure \ref{scitest.eps} for the result of
a numerical test.

\subsection{M5 from M2}
Let us apply our method to the theory realized on $N$ coincident M5-branes,
which we denote by $T=M5$.
It is the $6$-dim ${\cal N}=(2,0)$ theory of
$A_{N-1}$ type together with a free tensor multiplet.

The superconformal index is defined by \cite{Bhattacharya:2008zy}
\begin{align}
I_N^{(M5)}
=\tr[(-1)^F
\check q_1^{\check J_{12}}
\check q_2^{\check J_{34}}
\check q_3^{\check J_{56}}
\check x_1^{\check R_{12}}
\check x_2^{\check R_{34}}]\quad
\check q_1\check q_2\check q_3
=\check x_1\check x_2,
\label{m5index}
\end{align}
where $\check J_{12}$, $\check J_{34}$, and $\check J_{56}$
are Cartan generators of $SO(6)_{\rm spin}$
and $\check R_{12}$ and $\check R_{34}$ are
Cartan generators of $SO(5)_R$.

The index is given as a double giant graviton expansion
associated with two two-cycles in \cite{Arai:2020uwd},
and the single giant graviton sector has been studied.
The cycles are defined as the fixed loci of
$\check R_{12}$ and $\check R_{34}$ in $M_{M5}=S^4$.
We call $\check R_{12}$ ($\check R_{34}$) fixed locus ``12-cycle'' (``34-cycle'').
M2-branes wrapped around these cycles contribute to the index.

The superconformal index of the $3$d ${\cal N}=8$ SCFT
realized on coincident M2-branes
is defined by \cite{Bhattacharya:2008zy}
\begin{align}
I_N^{(M2)}
=\tr[(-1)^F
\hat q^{\hat J_{12}}
\hat x_1^{\hat R_{12}}
\hat x_2^{\hat R_{34}}
\hat x_3^{\hat R_{56}}
\hat x_4^{\hat R_{78}}
]\quad
\hat q=
\hat x_1
\hat x_2
\hat x_3
\hat x_4,
\label{m2index}
\end{align}
where $\hat J_{12}$ is the
spin and
$\hat R_{12}$,
$\hat R_{34}$,
$\hat R_{56}$, and
$\hat R_{78}$ are
$SO(8)_R$ Cartan generators.
We can calculate this index for an arbitrary $N$
by applying the localization method \cite{Kim:2009wb}
to the ABJM theory with Chern-Simons level $k=1$ \cite{Aharony:2008ug}.

The Cartan generators of the $3$d ${\cal N}=8$ superconformal algebra
acting on M2-branes wrapped around $12$-cycle
and those of the $6$d ${\cal N}=(2,0)$ superconformal algebra
acting on the $AdS_7$ boundary are
related by \cite{Arai:2020uwd}
\begin{align}
(\hat H,\hat J_{12},\hat R_{12},\hat R_{34},\hat R_{56},\hat R_{78})
=(\tfrac{1}{2}\check H-\tfrac{3}{2}\check R_{12},
\check R_{34},\check J_{12},\check J_{34},\check J_{56},-\check R_{12}).
\end{align}
Correspondingly, the fugacities in (\ref{m5index}) and those in (\ref{m2index}) are related by
\begin{align}
\sigma_{12}(\hat q,\hat x_1,\hat x_2,\hat x_3,\hat x_4)
=(\check x_2,\check q_1,\check q_2,\check q_3,\check x_1^{-1}).
\label{sigma12}
\end{align}
The variable change associated with $34$-cycle,
$\sigma_{34}$,
is given by swapping $\check x_1$ and $\check x_2$ after $\sigma_{12}$.
the M2-giant contributions $F_{12,N}^{(M5)}$
are obtained from the index $I_N^{(M2)}$ by
the relation
\begin{align}
F_{12,N}^{(M5)}=\sigma_{12}I_N^{(M2)}.
\label{f12andi}
\end{align}

Again, there are four independent variables and the lattice is four-dimensional.
We want to define a projection to two-dimensional lattice to show
an expansion as a two-dimensional plot.
We introduce variables $x$, $y$, and $u_i$ ($i=1,2,3$) by
\begin{align}
\check q_i=yu_i\quad
(u_1u_2u_3=1),\quad
\check x_1=x,\quad
\check x_2=x^{-1}y^3.
\label{m5unref}
\end{align}
We focus on $x$ and $y$ to show series in figures.

We also rewrite the fugacities for M2-branes as follows:
\begin{align}
\hat q=xy^3,\quad
\hat x_i=yu_i\quad(u_1u_2u_3=1),\quad
\hat x_4=x.
\end{align}
Then, the variable change
(\ref{sigma12}) becomes
\begin{align}
\sigma_{12}(x,y,u_i)
=(x^{-1},y,u_i).
\label{sigma12b}
\end{align}
Namely, (\ref{f12andi}) becomes
\begin{align}
F_{12,N}^{(M5)}(x,y,u_i)=I_N^{(M2)}(x^{-1},y,u_i).
\label{f12andi2}
\end{align}

\begin{figure}[htb]
\centering
\includegraphics{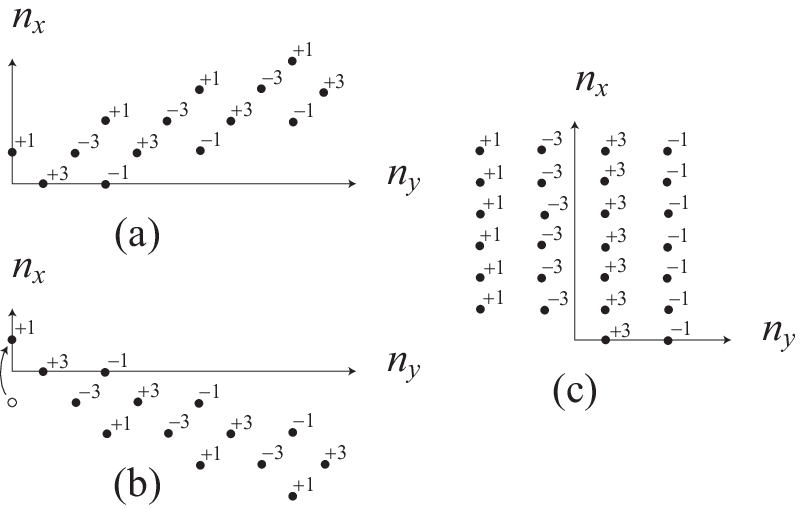}
\caption{(a) The letter index $i_1^{(M2)}$.
(b) The modification of $\sigma_1i_1^{(M2)}$ for $R_{yx}$.
(c) Infinitely many dots of $\sigma_2i_1^{(M2)}$ are out of $R_{yx}$.}\label{m2letters.eps}
\end{figure}

As in the previous examples,
let us first look at the letter index for a single giant graviton.
The theory on a single M2-brane is the free theory of an
${\cal N}=8$ scalar multiplet with the letter index \cite{Bhattacharya:2008zy}
\begin{align}
i_1^{(M2)}
=\frac{\hat x_1+\hat x_2+\hat x_3+\hat x_4
-\hat q(\hat x_1^{-1}+\hat x_2^{-1}+\hat x_3^{-1}+\hat x_4^{-1})}{1-\hat q}.
\end{align}
We show $i_1^{(M2)}$,
$\sigma_{12}i_1^{(M2)}$, and $\sigma_{34}i_1^{(M2)}$
projected on the two-dimensional plane
in Figure \ref{m2letters.eps}.
If we take the domain $R_{yx}$ defined in (\ref{hyx}),
$\sigma_{34}i_1^{(M2)}$
has infinitely many points outside $R_{yx}$.
They give the factor
\begin{align}
P=\prod_{k=1}^\infty(-y^3x^{-k})\prod_{k=1}^\infty(-yx^{-k})^{-3}.
\end{align}
If we take the product for each $k$ first, we obtain
$\prod_{k=1}^\infty x^{2k}=x^{\infty}$ and the contribution decouples.
Just like the previous examples,
we simply assume that $F_{m_{12},m_{34}}^{(M5)}$ with $m_{34}\geq1$ decouple,
and let us consider the simple expansion
associated with the $12$-cycle
\begin{align}
\frac{I_N^{(M5)}}{I_\infty^{(M5)}}
=\sum_{m=0}^\infty x^{mN}F_{12,m}^{(M5)}.
\label{m5expansion}
\end{align}

Let us numerically confirm (\ref{m5expansion}) holds for small $N$.
$I_N^{(M5)}$ with $N=-1,0,1$ and $\infty$ are given by
\begin{align}
I_{-1}^{(M5)}=0,\quad
I_0^{(M5)}=1,\quad
I_1^{(M5)}=\Pexp i_1^{(M5)},\quad
I_\infty^{(M5)}=\Pexp i_\infty^{(M5)},
\end{align}
where $i_1^{(M5)}$ and $i_\infty^{(M5)}$ are
the letter indices of the tensor multiplet and
the supergravity multiplet in $AdS_7\times S^4$.
They are given by \cite{Bhattacharya:2008zy}
\begin{align}
i_1^{(M5)}=\frac{
\check x_1+\check x_2
-\check q_1\check q_2
-\check q_2\check q_3
-\check q_3\check q_1
+\check x_1\check x_2}
{
(1-\check q_1)
(1-\check q_2)
(1-\check q_3)},
\end{align}
and
\begin{align}
i_\infty^{(M5)}
=\frac{
\check x_1+\check x_2
-\check q_1\check q_2
-\check q_2\check q_3
-\check q_3\check q_1
+\check x_1\check x_2
(
\check q_1
+\check q_2
+\check q_3
-\check x_1
-\check x_2)
}{(1-\check x_1)
(1-\check x_2)
(1-\check q_1)
(1-\check q_2)
(1-\check q_3)}.
\end{align}
By using these we can calculate the
left hand side of (\ref{m5expansion}).
In the following we calculate the right hand side of (\ref{m5expansion}).

\begin{figure}[htb]
\centering
\includegraphics{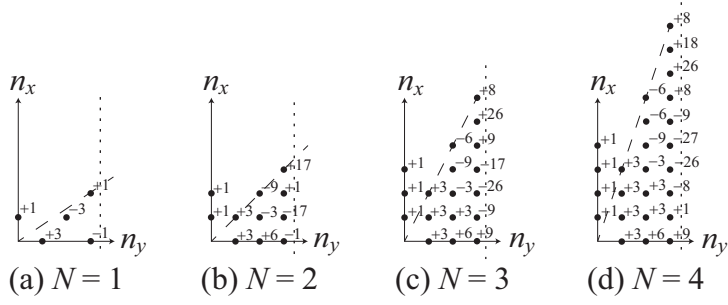}
\caption{$i_N^{(M2)}=\Plog I_N^{(M2)}$
for $N=1,2,3,4$ are shown.
As shown by the dotted lines the $y$ expansions are cut off at $n_y=3$.
}\label{abjm1.eps}
\end{figure}

\begin{figure}[htb]
\centering
\includegraphics{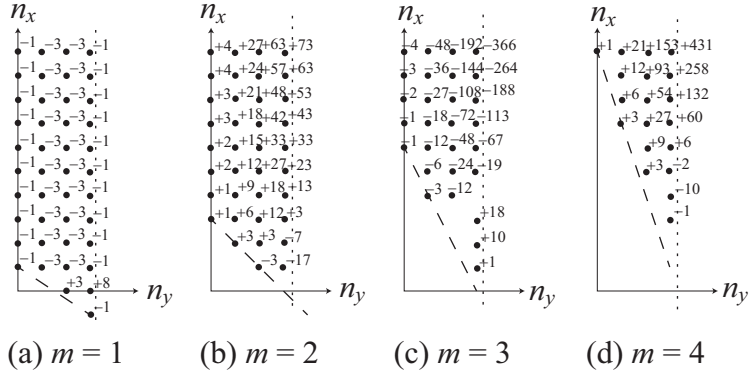}
\caption{M2-giant contributions $F_{12,m}^{(M5)}$ with
wrapping number $m=1,2,3,4$ are shown.
As shown by the dotted lines the $y$ expansions are cut off at $n_y=3$.
}\label{m2gg.eps}
\end{figure}

We first calculate the index $I_N^{(M2)}$ for different $N$
with the physical domain $R_{yx}$.
This is done by applying localization method \cite{Kim:2009wb}
to the ABJM theory \cite{Aharony:2008ug}.
The plethystic logarithms $i_N^{(M2)}$ for small $N$ are shown in Figure \ref{abjm1.eps}.
They have similar structure to $i_N^{(D3)}$.
In particular, for each $N$ there are
$N$ dots
on the positive part of the vertical axis
corresponding to $N$ $1/2$ BPS operators.
After the variable change (\ref{sigma12b}),
these dots have to be moved to the opposite points
to obtain the modified letter index.
The expansion of the analytically continued plethystic exponential
$F_{12,N}^{(M5)}|_{R_{yx}}=\Pexp_{R_{yx}}\sigma_{12}i_N^{(M2)}$ for $N=1,2,3,4$ are
shown in Figure \ref{m2gg.eps}.
For $N\geq2$ the dots are in the region above the
line $n_x=\frac{m(m+1)}{2}+mN-mn_y$.
($N=1$ case is exceptional and the line is $n_x=1-\frac{2}{3}n_y$.)

\begin{figure}[htb]
\centering
\includegraphics{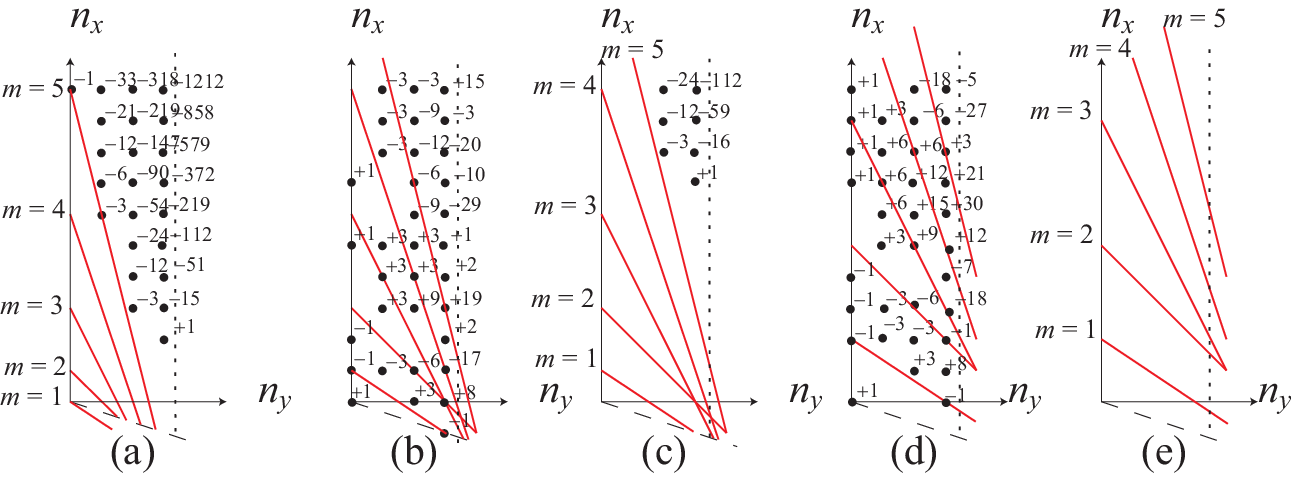}
\caption{The results of consistency check for $N=-1,0,1$.
$I_N^{(M5)}/I_\infty^{(M5)}$ for $N=0$ and $N=1$ are shown in (b) and (d),
respectively. Trivial one, $I_{-1}^{(M5)}/I_\infty^{(M5)}=0$ is not shown.
$(I_N^{(M5)}/I_\infty^{(M5)})-\sum_{m=0}^4x^{mN}F_{12,m}^{(M5)}$ for $N=-1$, $0$, and $1$ are shown in
(a), (c), and (e),respectively.
As shown by the dotted lines the $y$ expansions are cut off at $n_y=3$.
In (a), (c), and (e), all terms below the $m=5$ lines are correctly canceled,
and only expected errors due to $m\geq5$ contributions are left.
}\label{m5check.eps}
\end{figure}

With the functions $F_{12,m}^{(M5)}$ obtained above
we can confirm that (\ref{m5expansion}) holds
for $N=-1,0,1$ up to expected errors.
See Figure \ref{m5check.eps}.
This strongly suggests that
(\ref{m5expansion}) correctly gives
$I_N^{(M5)}$ for an arbitrary $N$.
Although we only show the two-dimensional plots,
we can calculate the full superconformal index.
See Appendix \ref{full.sec}.

We can also compare our results
with the analytic result for the Schur-like limit \cite{Kim:2013nva,Beem:2014kka}
obtained from the analysis of
five-dimensional SYM.
It is known that by setting $\check q_1=\check x_1$,
the index $I_N^{(M5)}$ reduces
to the following function of the single variable $\check x_1$:
\begin{align}
I_N^{(M5)}=\Pexp\frac{\check x_1+\check x_1^2+\cdots+\check x_1^N}{1-\check x_1}.
\label{schurlike}
\end{align}
If we take the same limit in our results shown in Appendix \ref{full.sec}
we obtain
\begin{align}
I_2^{(M5)}&=1+\check x_1+3\check x_1^2+5\check x_1^3+10\check x_1^4+\underline{15\check x_1^5}+{\cal O}(\check x_1^6),\nonumber\\
I_3^{(M5)}&=1+\check x_1+3\check x_1^2+6\check x_1^3+12\check x_1^4+\underline{20\check x_1^5}+{\cal O}(\check x_1^6),\nonumber\\
I_4^{(M5)}&=1+\check x_1+3\check x_1^2+6\check x_1^3+13\check x_1^4+\underline{22\check x_1^5}+{\cal O}(\check x_1^6),\nonumber\\
I_5^{(M5)}&=1+\check x_1+3\check x_1^2+6\check x_1^3+13\check x_1^4+\underline{23\check x_1^5}+{\cal O}(\check x_1^6).
\label{im52345}
\end{align}
We only showed terms independent of $\check q_2$ and $\check q_3$.
The underlines indicate the terms are incorrect.
These are consistent with (\ref{schurlike})
because in our numerical results the $y$ expansion is cut off at $n_y=3$ and
the expected order of errors
is ${\cal O}(\check x_1^4)$ or higher.
The $\check q_2$ or $\check q_3$-dependent terms which we did not show in (\ref{im52345})
are also consistent with the expected errors.

Note that the Schur-like limit is ill-defined for each M2-giant contribution
because
the limit corresponds to $\hat x_1=\hat x_4^{-1}$ via (\ref{sigma12}).
The terms $\hat x_1$ and $\hat x_1^{-1}$
appear at the same time in the letter index $i_1^{(M2)}$ in the limit,
and there is no domain covering both of them.
We need to take the limit after summing up
the contributions of M2-giants.

\subsection{M2 from M5}\label{m2fromm5.sec}
It is possible to interchange
the roles of M2 and M5 in the previous subsection.
Namely, we can calculate the finite $N$ index of M2 theory
by summing up M5-giant contributions
by the relation
\begin{align}
\frac{I_N^{(M2)}|_{R_{yx}}}{I_\infty^{(M2)}|_{R_{yx}}}
=\sum_{m=0}^\infty x^{mN}(\sigma_{12}^{-1}I_m^{(M5)})|_{R_{yx}},
\label{m2fromm5}
\end{align}
where the large $N$ index is given by $I_\infty^{(M2)}=\Pexp i_\infty^{(M2)}$ with
the letter index \cite{Bhattacharya:2008zy}
\begin{align}
i_\infty^{(M2)}
=\frac{(1-\hat q\hat x_1^{-1})(1-\hat q\hat x_2^{-1})(1-\hat q\hat x_3^{-1})(1-\hat q\hat x_4^{-1})}{(1-\hat x_1)(1-\hat x_2)(1-\hat x_3)(1-\hat x_4)(1-\hat q)^2}-\frac{1-\hat q+\hat q^2}{(1-\hat q)^2}.
\end{align}
The calculation to test (\ref{m2fromm5}) is parallel to the previous example,
and we only give a brief explanation.

In \cite{Arai:2020uwd} four five-cycles
in $M_{M5}=S^7$ were taken into account.
As the previous examples
we can show for a single M5 giant that
if we take $R_{yx}$ only one of the four cycles, $\hat R_{78}$ fixed locus,
gives non-trivial contribution.
Hence, we focus on $F_{78,N}^{(M2)}$ associated with the cycle.

\begin{figure}[htb]
\centering
\includegraphics{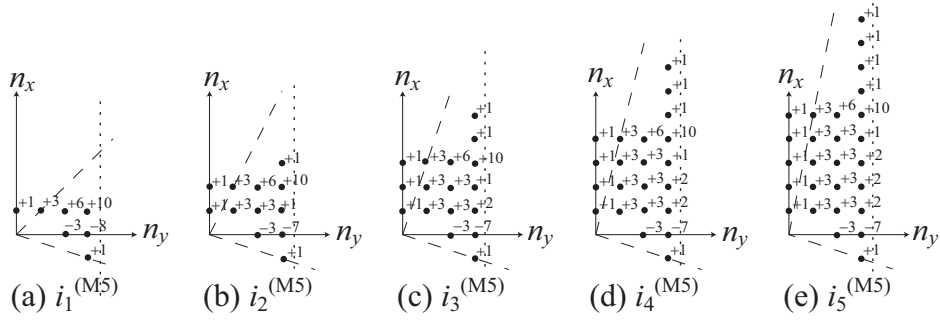}
\caption{$i_N^{(M5)}$ for $N=1,2,3,4,5$ are shown.
As shown by the dotted lines the $y$ expansions are cut off at $n_y=3$.
}\label{plogm5.eps}
\end{figure}

\begin{figure}[htb]
\centering
\includegraphics{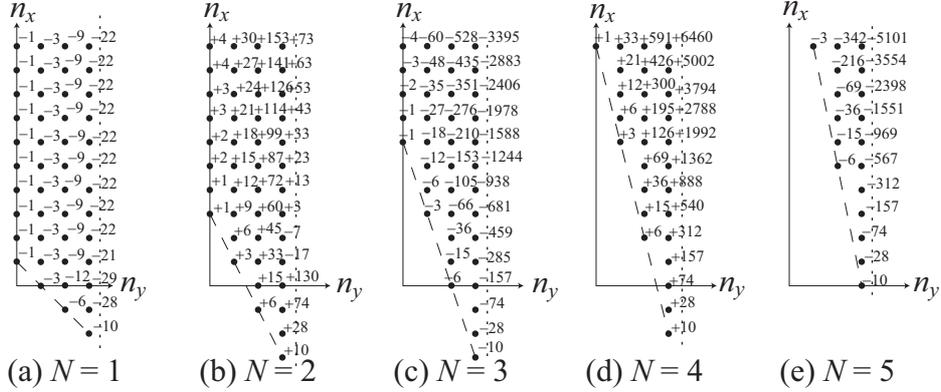}
\caption{M5 giants contributions $F_{78,N}^{(M2)}$ for $N=1,2,3,4,5$ are shown.
As shown by the dotted lines the $y$ expansions are cut off at $n_y=3$.
}\label{m5gg.eps}
\end{figure}

\begin{figure}[htb]
\centering
\includegraphics{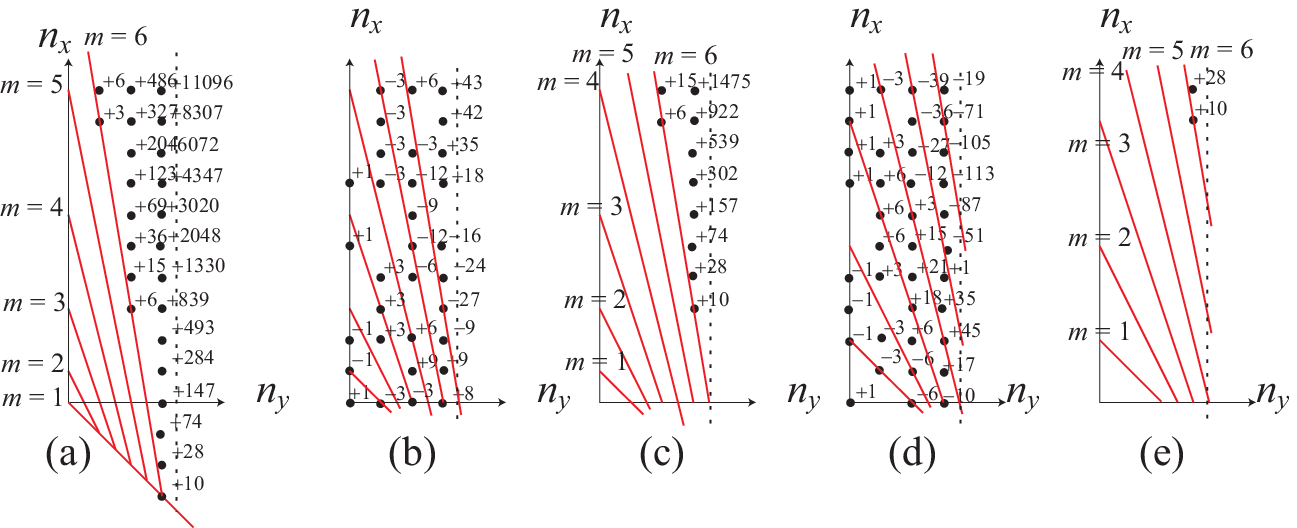}
\caption{
$(I_N^{(M2)}/I_\infty^{(M2)})$ for $N=0,1$ are shown
in (b) and (d), respectively.
The trivial one $I_{-1}^{(M2)}/I_\infty^{(M2)}=0$ is not shown.
$(I_N^{(M2)}/I_\infty^{(M2)})-\sum_{m=0}^5x^{mN}F_{78,m}^{(M2)}$ for $N=-1,0,1$ are
shown in (a), (c), and (e), respectively.
As shown by the dotted lines the $y$ expansions are cut off at $n_y=3$.
In (a), (c), and (e) all terms below $m=6$ lines are correctly canceled,
and only expected errors due to $m\geq6$ contributions are left.
}\label{m2check.eps}
\end{figure}

By using the results of previous subsection for $I_N^{(M5)}$
we can obtain the plethystic logarithms $i_N^{(M5)}=\Plog I_N^{(M5)}$
for small $N$ shown in Figure \ref{plogm5.eps}.
As the previous examples
we find $N$ dots
associated with
the $1/2$ BPS operators on the positive part of the
vertical axis for each $N$.
The functions $F_{78,N}^{(M2)}=\Pexp_{R_{yx}}\sigma_{12}^{-1}i_N^{(M5)}$ are shown
in Figure \ref{m5gg.eps}.
The results of a numerical check of (\ref{m2fromm5})
for $N=-1,0,1$ are
shown in Figure \ref{m2check.eps}.

\section{Conclusions and Discussion}
We investigated the giant graviton expansions.
In particular,
we concentrated on coincident giant gravitons wrapped around
a single cycle.
Because the expansion domain for giant gravitons and
that of the boundary theory are different,
we need to perform an analytic continuation
to relate them.
We proposed a simple prescription
(\ref{proposal}) to realize it.

We explicitly showed for the Schur index of $T=D3$
that different choices of the expansion domains
give different functions for
single giant graviton contributions
$F_{i,1}^{(D3)}$ ($i=x,y$).
If we use $R_{qu}$ in (\ref{rqudef}) both
$F_{x,1}^{(D3)}$ and
$F_{y,1}^{(D3)}$ contribute to the Schur index,
while if we use $R_{yx}$ in (\ref{hyx}),
$F_{y,1}^{(D3)}$ does not contribute.
Although we have not proved $F_{y,N\geq2}^{(D3)}$ also vanish,
this partially explains why there are two different expansions:
the simple expansion found in \cite{Gaiotto:2021xce}
and the multiple expansion in \cite{Arai:2020qaj}.
It is surprising that they give the same result
even though some contributions are lost in the simple expansion.
This may be related to the fact that the set of functions
$F_{m_x,m_y}$ are strongly constrained.
For example, we can formally substitute negative $N$ to expansion (\ref{multipleexp})
or (\ref{simpleexp}), and find that the result is vanishing.
This gives very strong constraints on the functions.
This implies that the functions share common information,
and only small subset appearing in the simple expansion
may be sufficient to give the complete answer.
It would be interesting to investigate the structure of
the constrained set of the functions.

An important application of our method is
the calculation of the full superconformal index of
the 6d ${\cal N}=(2,0)$ $A_{N-1}$ SCFT.
It is in principle possible
to calculate $I_N^{(M5)}$ for an arbitrary $N$
up to an arbitrary order starting from
the indices $I_m^{(M2)}$ of the M2-brane theories with
different $m$.
Some results are shown in Appendix \ref{full.sec}.

An important merit of our method is that the theory on giant gravitons
does not have to be Lagrangian theories unlike
the method adopted in \cite{Arai:2020qaj,Imamura:2021ytr,Lee:2022vig}, which uses
deformed contours in the gauge fugacity integrals to realize the analytic
continuation.
We only need the final expression
of the index of the giant graviton theory.
It enables us to apply the method to M5-giants,
as was demonstrated in \ref{m2fromm5.sec}.
Another example with non-Lagrangian giant gravitons is $AdS_5\times S^5$ with
a $7$-brane insertion \cite{Fayyazuddin:1998fb,Aharony:1998xz}.
As is demonstrated in \cite{Imamura:2021dya},
the giant graviton expansion works well for this system
at least for the leading contribution.
There are higher order contributions coming from
giant gravitons coincident with the $7$-brane,
on which a non-Lagrangian theory is realized.
We expect our method is useful for the analysis of such contributions.

We studied two expansion domains $R_{yx}$ and $R_{qu}$ for the Schur index of $T=D3$.
An advantage of $R_{qu}$ over $R_{yx}$
is that for each value of $m=m_x+m_y$ the symmetry between $x$ and $y$ is manifest.
Unfortunately, it turned out that
our method has limited applicability for the calculation
with the domain $R_{qu}$
due to the ill-defined factor $P$.
It would be very nice if we can somehow
regularize the product in $P$ when it is infinite
and make it possible to apply the method to an arbitrary
contribution $F_{i,m}^{(T)}$.

In general, the functions $F_{m_1,\ldots,m_d}^{(T)}$
do not factorize into $F_{i,m}^{(T)}$.
In the case of D3-giants it is possible to write $F_{m_1,\ldots,m_d}^{(T)}$
in the matrix integral form.
Even so, it is not clear how we should choose the integration contours.
Because we cannot take a physical domain of expansion,
using unit circles for integration contours is not justified.
Although some rules for contours
have been proposed \cite{Imamura:2021ytr,Lee:2022vig},
it is desirable to find more efficient method of calculation applicable
to non-Lagrangian giant gravitons.

Another important problem is to clarify to what extent
the giant graviton expansion works.
In this work we focused only on the maximally supersymmetric theories:
4d ${\cal N}=4$ SYM on D3-branes,
3d ${\cal N}=8$ SCFT on M2-branes,
and 6d ${\cal N}=(2,0)$ SCFT on M5-branes.
The analysis on the gauge theory side \cite{Gaiotto:2021xce,Lee:2022vig}
found
the structure of the giant graviton expansions
in variety of theories.
It would be interesting to study the applicability of our method
to more general class of theories
which have holographic dual description.

In addition to the giant graviton expansions
(\ref{multipleexp}) and (\ref{simpleexp}),
another similar expansion was proposed in \cite{Murthy:2022ien}.
It may be interesting to study the relation among them.

\section*{Acknowledgments}
The author would like to thank S.~Murayama and D.~Yokoyama for valuable discussions.
The work of Y.~I. was
partially supported by Grand-in-Aid for Scientific Research (C) (No.21K03569),
Ministry of Education, Science and Culture, Japan.
This work used computational resources TSUBAME3.0 supercomputer provided by Tokyo Institute of Technology.

\appendix
\section{Full index of $6$-dim $A_{N-1}$ SCFT}\label{full.sec}
Let $\chi_{[a,b]}$ be the $SU(3)$ character of the representation with
Dynkin label $[a,b]$ defined such that
$\chi_{[1,0]}=u_1+u_2+u_3$.
The superconformal indices $I_N^{(M5)}$ for $N=2,3,4,5$
are given as follows.
\newcommand{\x}{\check x_1}
\begin{align}
I^{(M5)}_2
&=(1+\x+2\x^2+2\x^3+3\x^4+3\x^5+4\x^6+4\x^7+5\x^8+5\x^9+6\x^{10}+\cdots)
\nonumber\\
&+y\chi_{[1,0]}(\x+2\x^2+3\x^3+4\x^4+5\x^5+6\x^6+7\x^7+8\x^8+9\x^9+10\x^{10}+\cdots)
\nonumber\\
&+y^2[\chi_{[0,1]}(-1-2\x-3\x^2-3\x^3-4\x^4-4\x^5-5\x^6-5\x^7-6\x^8-6\x^9-7\x^{10}+\cdots)
\nonumber\\
&\quad+\chi_{[2,0]}(\x+3\x^2+5\x^3+8\x^4+10\x^5+13\x^6+15\x^7+18\x^8+20\x^9+23\x^{10}+\cdots)]
\nonumber\\
&+y^3[(\x^{-1}+2+\x-\x^3-2\x^4-3\x^5-4\x^6-5\x^7-6\x^8-7\x^9-8\x^{10}+\cdots)
\nonumber\\
&\quad+\chi_{[1,1]}(-1-3\x-5\x^2-6\x^3-6\x^4-6\x^5-6\x^6-6\x^7-6\x^8-6\x^9-6\x^{10}+\cdots)
\nonumber\\
&\quad+\chi_{[3,0]}(\x+3\x^2+7\x^3+11\x^4+16\x^5+21\x^6+26\x^7+31\x^8+36\x^9+41\x^{10}+\cdots)]
\nonumber\\
&+\cdots
\end{align}
\begin{align}
I^{(M5)}_3
&=(1+\x+2\x^2+3\x^3+4\x^4+5\x^5+7\x^6+8\x^7+10\x^8+12\x^9+14\x^{10}+\cdots)
\nonumber\\
&+y\chi_{[1,0]}(\x+2\x^2+4\x^3+6\x^4+9\x^5+12\x^6+16\x^7+20\x^8+25\x^9+30\x^{10}+\cdots)
\nonumber\\
&+y^2[\chi_{[0,1]}(-1-2\x-4\x^2-5\x^3-7\x^4-8\x^5-10\x^6-11\x^7-13\x^8-14\x^9-16\x^{10}+\cdots)
\nonumber\\
&\quad+\chi_{[2,0]}(\x+3\x^2+6\x^3+11\x^4+17\x^5+25\x^6+34\x^7+45\x^8+57\x^9+71\x^{10}+\cdots)]
\nonumber\\
&+y^3[(\x^{-1}+2+2\x+\x^2-\x^3-4\x^4-8\x^5-12\x^6-18\x^7-24\x^8-31\x^9-39\x^{10}+\cdots)
\nonumber\\
&\quad+\chi_{[1,1]}(-1-3\x-6\x^2-10\x^3-13\x^4-16\x^5-18\x^6-19\x^7-19\x^8-19\x^9-17\x^{10}+\cdots)
\nonumber\\
&\quad+\chi_{[3,0]}(\x+3\x^2+8\x^3+15\x^4+26\x^5+40\x^6+58\x^7+79\x^8+105\x^9+133\x^{10}+\cdots)]
\nonumber\\
&+\cdots
\end{align}
\begin{align}
I^{(M5)}_4
&=(1+\x+2\x^2+3\x^3+5\x^4+6\x^5+9\x^6+11\x^7+15\x^8+18\x^9+23\x^{10}+\cdots)
\nonumber\\
&+y\chi_{[1,0]}(\x+2\x^2+4\x^3+7\x^4+11\x^5+16\x^6+23\x^7+31\x^8+41\x^9+53\x^{10}+\cdots)
\nonumber\\
&+y^2[\chi_{[0,1]}(-1-2\x-4\x^2-6\x^3-9\x^4-11\x^5-15\x^6-17\x^7-21\x^8-23\x^9-27\x^{10}+\cdots)
\nonumber\\
&\quad+\chi_{[2,0]}(\x+3\x^2+6\x^3+12\x^4+20\x^5+32\x^6+47\x^7+68\x^8+92\x^9+124\x^{10}+\cdots)]
\nonumber\\
&+y^3[(\x^{-1}+2+2\x+2\x^2-4\x^4-10\x^5-18\x^6-30\x^7-44\x^8-62\x^9-84\x^{10}+\cdots)
\nonumber\\
&\quad+\chi_{[1,1]}(-1-3\x-6\x^2-11\x^3-17\x^4-23\x^5-30\x^6-36\x^7-40\x^8-43\x^9-43\x^{10}+\cdots)
\nonumber\\
&\quad+\chi_{[3,0]}(\x+3\x^2+8\x^3+16\x^4+30\x^5+50\x^6+79\x^7+117\x^8+168\x^9+231\x^{10}+\cdots)]
\nonumber\\
&+\cdots
\end{align}
\begin{align}
I^{(M5)}_5
&=(1+\x+2\x^2+3\x^3+5\x^4+7\x^5+10\x^6+13\x^7+18\x^8+23\x^9+30\x^{10}+\cdots)
\nonumber\\
&+y\chi_{[1,0]}(\x+2\x^2+4\x^3+7\x^4+12\x^5+18\x^6+27\x^7+38\x^8+53\x^9+71\x^{10}+\cdots)
\nonumber\\
&+y^2[\chi_{[0,1]}(-1-2\x-4\x^2-6\x^3-10\x^4-13\x^5-18\x^6-22\x^7-28\x^8-32\x^9-38\x^{10}+\cdots)
\nonumber\\
&\quad+\chi_{[2,0]}(\x+3\x^2+6\x^3+12\x^4+21\x^5+35\x^6+54\x^7+81\x^8+116\x^9+163\x^{10}+\cdots)]
\nonumber\\
&+y^3[(\x^{-1}+2+2\x+2\x^2+\x^3-3\x^4-10\x^5-20\x^6-36\x^7-57\x^8-86\x^9-123\x^{10}+\cdots)
\nonumber\\
&\quad+\chi_{[1,1]}(-1-3\x-6\x^2-11\x^3-18\x^4-27\x^5-37\x^6-48\x^7-59\x^8-69\x^9-75\x^{10}+\cdots)
\nonumber\\
&\quad+\chi_{[3,0]}(\x+3\x^2+8\x^3+16\x^4+31\x^5+54\x^6+89\x^7+138\x^8+208\x^9+300\x^{10}+\cdots)]
\nonumber\\
&+\cdots
\end{align}

We can also calculate the index of $A_{N-1}$ SCFT $I_N^{(A)}$
by removing the contribution of the free tensor multiplet by
\begin{align}
I_N^{(A)}=\frac{I_N^{(M5)}}{I_1^{(M5)}}.
\end{align}
The numerical results for $N=2,3,4,5$ are as follows:
\begin{align}
I^{(A)}_2
&=(1+\x^2+\x^4+\x^6+\x^8+\x^{10}+\cdots)
\nonumber\\
&+y\chi_{[1,0]}(\x^2+\x^4+\x^6+\x^8+\x^{10}+\cdots)
\nonumber\\
&+y^2[\chi_{[0,1]}(-\x-\x^3-\x^5-\x^7-\x^9+\cdots)
\nonumber\\
&\quad+\chi_{[2,0]}(\x^2+2\x^4+2\x^6+2\x^8+2\x^{10}+\cdots)]
\nonumber\\
&+y^3[(1-\x+\x^2-x^3+\x^4-\x^5+\x^6-\x^7+\x^8-\x^9+\x^{10}+\cdots)
\nonumber\\
&\quad+\chi_{[1,1]}(-\x-2\x^3+\x^4-2\x^5+\x^6-2\x^7+\x^8-2\x^9+\x^{10}+\cdots)
\nonumber\\
&\quad+\chi_{[3,0]}(\x^2+2\x^4+3\x^6+3\x^8+3\x^{10}+\cdots)]
\nonumber\\
&+\cdots
\end{align}
\begin{align}
I^{(A)}_3
&=(1+\x^2+\x^3+\x^4+\x^5+2\x^6+\x^7+2\x^8+2\x^9+2\x^{10}+\cdots)
\nonumber\\
&+y\chi_{[1,0]}(\x^2+\x^3+\x^4+2\x^5+2\x^6+2\x^7+3\x^8+3\x^9+3\x^{10}+\cdots)
\nonumber\\
&+y^2[\chi_{[0,1]}(-\x-\x^2-\x^3-2\x^4-\x^5-2\x^6-2\x^7-2\x^8-2\x^9-3\x^{10}+\cdots)
\nonumber\\
&\quad+\chi_{[2,0]}(\x^2+\x^3+2\x^4+3\x^5+4\x^6+4\x^7+6\x^8+6\x^9+7\x^{10}+\cdots)]
\nonumber\\
&+y^3[(1-\x^4-\x^5-\x^6-2\x^7-2\x^8-2\x^9-3\x^{10}+\cdots)
\nonumber\\
&\quad+\chi_{[1,1]}(-\x-\x^2-2\x^3-3\x^4-2\x^5-3\x^6-3\x^7-2\x^8-3\x^9-3\x^{10}+\cdots)
\nonumber\\
&\quad+\chi_{[3,0]}(\x^2+\x^3+2\x^4+4\x^5+5\x^6+6\x^7+9\x^8+10\x^9+11\x^{10}+\cdots)]
\nonumber\\
&+\cdots
\end{align}
\begin{align}
I^{(A)}_4
&=(1+\x^2+\x^3+2\x^4+\x^5+3\x^6+2\x^7+4\x^8+3\x^9+5\x^{10}+\cdots)
\nonumber\\
&+y\chi_{[1,0]}(\x^2+\x^3+2\x^4+2\x^5+4\x^6+4\x^7+6\x^8+6\x^9+9\x^{10}+\cdots)
\nonumber\\
&+y^2[\chi_{[0,1]}(-\x-\x^2-2\x^3-2\x^4-3\x^5-3\x^6-4\x^7-4\x^8-5\x^9-5\x^{10}+\cdots)
\nonumber\\
&\quad+\chi_{[2,0]}(\x^2+\x^3+3\x^4+3\x^5+7\x^6+7\x^7+12\x^8+12\x^9+19\x^{10}+\cdots)]
\nonumber\\
&+y^3[(1+\x^2-\x^3-3\x^5-2\x^6-6\x^7-5\x^8-9\x^9-9\x^{10}+\cdots)
\nonumber\\
&\quad+\chi_{[1,1]}(-\x-\x^2-3\x^3-3\x^4-6\x^5-5\x^6-8\x^7-6\x^8-10\x^9-6\x^{10}+\cdots)
\nonumber\\
&\quad+\chi_{[3,0]}(\x^2+\x^3+3\x^4+4\x^5+9\x^6+10\x^7+17\x^8+20\x^9+30\x^{10}+\cdots)]
\nonumber\\
&+\cdots
\end{align}
\begin{align}
I^{(A)}_5
&=(1+\x^2+\x^3+2\x^4+2\x^5+3\x^6+3\x^7+5\x^8+5\x^9+7\x^{10}+\cdots)
\nonumber\\
&+y\chi_{[1,0]}(\x^2+\x^3+2\x^4+3\x^5+4\x^6+6\x^7+8\x^8+10\x^9+13\x^{10}+\cdots)
\nonumber\\
&+y^2[\chi_{[0,1]}(-\x-\x^2-2\x^3-3\x^4-3\x^5-5\x^6-5\x^7-7\x^8-7\x^9-9\x^{10}+\cdots)
\nonumber\\
&\quad+\chi_{[2,0]}(\x^2+\x^3+3\x^4+4\x^5+7\x^6+10\x^7+15\x^8+19\x^9+27\x^{10}+\cdots)]
\nonumber\\
&+y^3[(1+\x^2-\x^4-2\x^5-4\x^6-7\x^7-10\x^8-14\x^9-19\x^{10}+\cdots)
\nonumber\\
&\quad+\chi_{[1,1]}(-\x-\x^2-3\x^3-4\x^4-6\x^5-9\x^6-10\x^7-13\x^8-15\x^9-16\x^{10}+\cdots)
\nonumber\\
&\quad+\chi_{[3,0]}(\x^2+\x^3+3\x^4+5\x^5+9\x^6+14\x^7+21\x^8+30\x^9+42\x^{10}+\cdots)]
\nonumber\\
&+\cdots
\end{align}

\end{document}